\pdfoutput=1

\documentclass[11pt]{article}

\usepackage[final]{acl}

\usepackage{times}
\usepackage{latexsym}

\usepackage[T1]{fontenc}

\usepackage[utf8]{inputenc}

\usepackage{microtype}

\usepackage{inconsolata}

\usepackage{graphicx}

\usepackage{adjustbox,booktabs,multirow}
\usepackage{listings}
\usepackage{float}
\usepackage{enumitem}
\usepackage{multirow} 
\usepackage{array} 
\usepackage{graphicx}
\usepackage{arydshln}
\usepackage{tablefootnote}
\usepackage{url}

\newcommand{\derefine}{\textit{Refined-Decompile}}
\newcommand{\deend}{\textit{End2end-Decompile}}

\newcommand{\ghidra}{Ghidra}

\title{LLM4Decompile: Decompiling Binary Code with Large Language Models}

\author{
 \textbf{Hanzhuo Tan\textsuperscript{1,2}},
 \textbf{Qi Luo\textsuperscript{1}},
 \textbf{Jing Li\textsuperscript{2,3}},
 \textbf{Yuqun Zhang\textsuperscript{1}\thanks{Yuqun Zhang is the corresponding author.}},
\\
 \textsuperscript{1}Department of Computer Science and Engineering, \\ Southern University of Science and Technology, Shenzhen, China \\
 \textsuperscript{2}Department of Computing, \\ The Hong Kong Polytechnic University, HKSAR, China \\
 \textsuperscript{3} Research Centre for Data Science \& Artificial Intelligence\\
\\
 \small{
hanzhuo.tan@connect.polyu.hk, 12232440@mail.sustech.edu.cn, jing-amelia.li@polyu.edu.hk, zhangyq@sustech.edu.cn
 }
}

\begin{document}
\maketitle
\begin{abstract}

Decompilation aims to convert binary code to high-level source code, but traditional tools like Ghidra often produce results that are difficult to read and execute. Motivated by the advancements in Large Language Models (LLMs), we propose LLM4Decompile, the first and largest open-source LLM series (1.3B to 33B) trained to decompile binary code. 
We optimize the LLM training process and introduce the LLM4Decompile-End models to decompile binary directly. The resulting models significantly outperform GPT-4o and \ghidra{} on the HumanEval and ExeBench benchmarks over 100\% in terms of re-executability rate.
Additionally, we improve the standard refinement approach to fine-tune the LLM4Decompile-Ref models, enabling them to effectively refine the decompiled code from Ghidra and achieve a further 16.2\% improvement over the LLM4Decompile-End. LLM4Decompile\footnote{\url{https://github.com/albertan017/LLM4Decompile}} demonstrates the potential of LLMs to revolutionize binary code decompilation, delivering remarkable improvements in readability and executability while complementing conventional tools for optimal results.

\end{abstract}


\section{Introduction}
Decompilation, the reverse process of converting machine code or binary code into a high-level programming language, facilitates various reverse engineering tasks such as vulnerability identification, malware research, and legacy software migration~\cite{decompilation1, decompilation2_rnn, btc, lmpa, slade, nova,MLMZhang, refine_decompile,hu2024degpt}. Decompilation is challenging due to the loss of information inherent in the compilation process, particularly finer details such as variable names~\cite{variable_name} and fundamental structures like loops and conditionals~\cite{for_loop}. To address these challenges, numerous tools have been developed for decompilation, with Ghidra~\cite{ghidra} and IDA Pro~\cite{idapro} being the most commonly used. Although these tools have the capability to revert binary code to high-level pseudo-code, the outputs often lack readability and re-executability~\cite{decoompile_test, wang2017ramblr}, which are essential for applications like legacy software migration and security instrumentation tasks~\cite{refine_decompile,decompile_recompile}.

\begin{figure}[t]
  \centering
  \includegraphics[width=0.9\linewidth]{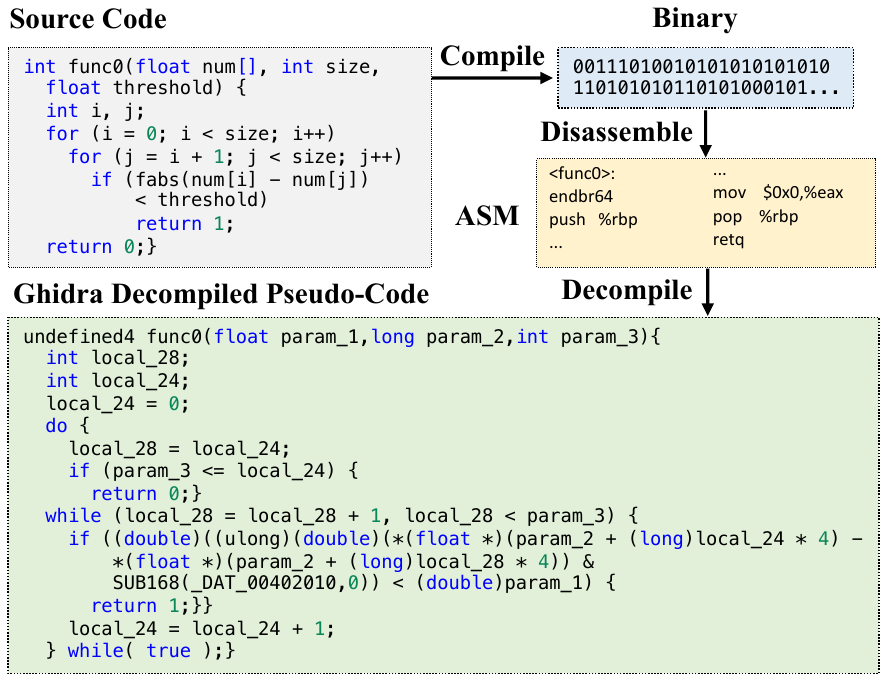}
  \caption{Illustration of compiling source code to binary, disassembling binary to assembly code (ASM), and decompiling ASM to pseudo-code with Ghidra. The pseudo-code is hard to read and not executable. }
  \label{fig:case}
  \vspace{-2em}
\end{figure}

Figure~\ref{fig:case} illustrates the transformation from the source C code to a binary file, assembly code (ASM), and pseudo-code decompiled from Ghidra. In this pseudo-code, the original nested \texttt{for} structure is replaced with a less intuitive combination of a \texttt{do-while} loop inside another \texttt{while} loop. Furthermore, array indexing like \texttt{num[i]} is decompiled into complicated pointer arithmetic such as \texttt{*(float *)(param\_2 + (long)local\_24 * 4)}. 
The decompiled output also exhibits syntactical errors, with the function return type being converted to \texttt{undefined4}. Overall, traditional decompilation tools often strip away the syntactic clarity provided by high-level languages and do not ensure the correctness of syntax, posing significant challenges even for skilled developers to reconstruct the algorithmic logic~\cite{refine_decompile,hu2024degpt}.

Recent advancements in Large Language Models (LLMs) have greatly improved the process of decompiling code. 
There are two primary approaches to LLM-based decompilation---\derefine{} and \deend{}. In particular, \derefine{} prompts LLMs to refine the results from traditional decompilation tools~\cite{hu2024degpt,refine_decompile,lmpa}. However, LLMs are primarily optimized for high-level programming languages and may not be as effective with binary data.
\deend{} fine-tunes LLMs to decompile binaries directly. Nevertheless, previous open-source applications of this approach were limited by the use of smaller models with only around 200 million parameters and restricted training corpus~\cite{btc,slade,nova}, In contrast, utilizing larger models trained on broader datasets has proven to substantially improve the performance~\cite{scalelaw1,scalelaw2,codellama,gpt4}.

To address the limitations of previous studies, we propose LLM4Decompile, the first and largest open-source LLM series with sizes ranging from 1.3B to 33B parameters specifically trained to decompile binary code. To the best of our knowledge, there's no previous study attempts to improve the capability of LLM-based decompilation in such depth or incorporate such large-scale LLMs. 
Based on the \deend{} approach, we introduce three critical steps: data augmentation, data cleaning, and two-stage training, to optimize the LLM training process and introduce the LLM4Decompile-End models to decompile binary directly.
Specifically, our LLM4Decompile-End-6.7B model demonstrates a successful decompilation rate of 45.4\% on HumanEval~\cite{humaneval} and 18.0\% on ExeBench~\cite{exebench}, far exceeding \ghidra{}~\cite{ghidra} or GPT-4o~\cite{gpt4} by over 100\%. 
Additionally, we improve the \derefine{} strategy by examining the efficiency of Ghidra's decompilation process, augmenting and filtering data to fine-tune the LLM4Decompile-Ref models, which excel at refining Ghidra's output.
Experiments suggest a higher performance ceiling for the enhanced \derefine{} approach, with 16.2\% improvement over LLM4Decompile-End. Additionally, we assess the risks associated with the potential misuse of our model under obfuscation conditions commonly used in software protection. Our findings indicate that neither our approach nor \ghidra{} can effectively decompile obfuscated code, mitigating concerns about unauthorized use for infringement of intellectual property.


In summary, our contributions are as follows:
\begin{itemize}[leftmargin=*, topsep=0pt]
    \item We introduce the LLM4Decompile series, the first and largest open-source LLMs (ranging from 1.3B to 33B parameters) fine-tuned on 15 billion tokens for decompilation. 
    \item We optimize the LLM training process and introduce LLM4Decompile-End models, which set a new performance standard of direct binary decompilation, significantly surpassing GPT-4o and \ghidra{} by over 100\% in terms of re-executability on the HumanEval and ExeBench benchmarks.
    \item We improve the \derefine{} approach to fine-tune the LLM4Decompile-Ref models, enabling them to effectively refine the decompiled results from Ghidra and achieve further 16.2\% re-executability enhancements over LLM4Decompile-End.
\end{itemize}

\section{Related Work}
The practice of reversing executable binaries to their source code form, known as decompilation, has been researched for decades~\cite{decompile_his1,decompile_his2,decompile_his3}. Traditional decompilation relies on analyzing the control and data flows of program~\cite{decompilation1}, and employing pattern matching, as seen in tools like Hex-Rays Ida pro~\cite{idapro} and \ghidra{}~\cite{ghidra}. These systems attempt to identify patterns within a program’s control-flow graph (CFG) that corresponding to standard programming constructs such as conditional statements or loops. However, the output from such decompilation processes tends to be a source-code-like representation of assembly code, including direct translations of variables to registers, use of gotos, and other low-level operations instead of the original high-level language constructs. This output, while often functionally similar to the original code, is difficult to understand and may not be re-executable~\cite{ioacc2,refine_decompile}.
Drawing inspiration from neural machine translation, researchers have reformulated decompilation as a translation exercise, converting machine-level instructions into readable source code~\cite{decompile_his3}. Initial attempts in this area utilized recurrent neural networks (RNNs)~\cite{decompilation2_rnn} for decompilation, complemented by error-correction techniques to enhance the outcomes. 


Motivated by the success of Large Language Models~\cite{starcoder,codellama,deepseekcoder}, researchers have employed LLMs for decompilation, primarily through two approaches---\derefine{} and \deend{}. In particular, \derefine{} prompts the LLMs to refine results from traditional decompilation tools like \ghidra{} or IDA Pro. For instance, DeGPT~\cite{hu2024degpt} enhances Ghidra's readability by reducing cognitive load by 24.4\%, while DecGPT~\cite{refine_decompile} increases IDA Pro's re-executability rate to over 75\% by integrating error messages into its refinement process. These approaches, however, largely ignore the fact that LLMs are designed primarily for high-level programming languages~\cite{starcoder,codellama,deepseekcoder}, and their effectiveness with binary files is not well-established. 
\deend{}, on the other hand, fine-tunes LLMs to decompile binaries directly. Early open-source models like BTC~\cite{btc} and recent development Slade~\cite{slade} adopt the language model with around 200 million parameters~\cite{lewis-etal-2020-bart} to fine-tune for decompilation. While Nova~\cite{nova}, which is not open-sourced, develops a binary LLM with 1 billion parameters and fine-tunes it for decompilation. Consequently, the largest open-source model in this domain is limited to 200M. Whereas utilizing larger models trained on broader datasets has proven to substantially improve the performance~\cite{scalelaw1,scalelaw2,codellama}.

Therefore, our objective is to present the first and most extensive open-source LLM4Decompile series, aiming at comprehensively advancing the decompilation capability of LLMs. Initially, we optimize the \deend{} approach to train the LLM4Decompile-End, demonstrating its effectiveness in directly decompiling binary files. Subsequently, we enhance the \derefine{} frameworks to integrate LLMs with Ghidra, augmenting traditional tools for optimal effectiveness.

\section{LLM4Decompile}
First, we introduce our strategy for optimizing LLM training to directly decompile binaries, the resulting models are named as LLM4Decompile-End. Following this, we detail our efforts for enhancing the \derefine{} approach, the corresponding fine-tuned models are referred to as LLM4Decompile-Ref, which can effectively refine the decompiled results from Ghidra.

\subsection{LLM4Decompile-End}\label{sec:end}
In this section, we describe the general \deend{} framework, and present details on our strategy to optimize the training of LLM4Decompile-End models.

\subsubsection{The End2End-Decompile Framework}

\begin{figure}[t]
  \centering
  \includegraphics[width=0.9\linewidth]{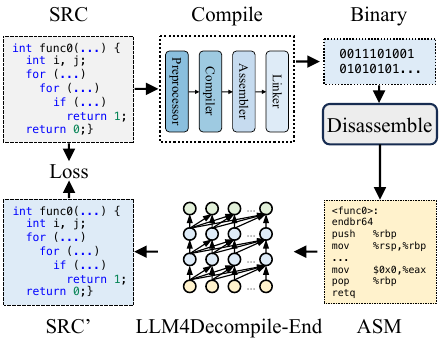}
  \caption{\deend{} framework. The source code (SRC) is compiled to binary, disassembled to assembly instructions (ASM), and decompiled by LLM4Decompile to generate SRC'. Loss is computed between SRC and SRC' for training.}
  \label{fig:compile}
\end{figure}

Figure~\ref{fig:compile} illustrates the \deend{} framework from compilation to decompilation processes. During compilation, the Preprocessor processes the source code (SRC) to eliminate comments and expand macros or includes. The cleaned code is then forwarded to the Compiler, which converts it into assembly code (ASM). This ASM is transformed into binary code (0s and 1s) by the Assembler. The Linker finalizes the process by linking function calls to create an executable file. Decompilation, on the other hand, involves converting binary code back into a source file. LLMs, being trained on text, lack the ability to process binary data directly. Therefore, binaries must be disassembled by \texttt{Objdump} into assembly language (ASM) first. It should be noted that binary and disassembled ASM are equivalent, they can be interconverted, and thus we refer to them interchangeably. Finally, the loss is computed between the decompiled code and source code to guide the training.

\subsubsection{Optimize LLM4Decompile-End}\label{sec:opt_end}
We optimize the training of LLM4Decompile-End Models through three key steps: 1) augmenting the training corpus, 2) improving the quality of the data, 3) and incorporating two-state training.

\paragraph{\textbf{Training Corpus.}} As indicated by the Scaling-Law~\cite{scalelaw1,scalelaw2}, the effectiveness of an LLM heavily relies on the size of the training corpus. Consequently, our initial step in training optimization involves incorporating a large training corpus. We construct asm-source pairs based on ExeBench~\cite{exebench}, which is the largest public collection of five million C functions. To further expand the training data, we consider the compilation optimization states frequently used by developers. 
The compilation optimization involves techniques like eliminating redundant instructions, better register allocation, and loop transformations~\cite{compiler_opt}, which perfectly acts as data augmentation for decompilation. The key optimization levels are O0 (default, no optimization) to O3 (aggressive optimizations). We compile the source code into all four stages, i.e., O0, O1, O2, and O3, and pair each of them with the source code. 

\paragraph{\textbf{Data Quality.}}
Data quality is critical in training an effective model~\cite{starcoder}. 
Therefore, our second step is to clean our training set. We follow the guidelines of StarCoder~\cite{starcoder} by computing MinHash~\cite{minhash} for the code and utilizing Locally Sensitive Hashing (LSH) to remove duplicates. We also exclude samples that are less than 10 tokens. 

\paragraph{\textbf{Two-Stage Training.}}\label{sec:two-stage} 
Our final step for training optimization aims to educate the model with binary knowledge, and includes two-stage training. In the first stage, we train the model with a large corpus of compilable but not linkable (executable) data.
Note that it's significantly easier to extract C code that is compilable but not linkable~\cite{anghabench,exebench}. Such not-executable binary object code will closely resemble its executable version except it lacks linked addresses for external symbols. 
Therefore, in the first stage, we use the extensive compilable codes to ground our model in binary knowledge. In the second stage, we refine the model using executable code to ensure its practical applicability. We also conduct an ablation study for the two-stage training in Section~\ref{sec:ablation}. Comparison between compilable and executable data is detailed in Appendix~\ref{sec:appen_exe}.


\subsection{LLM4Decompile-Ref}
We now examine how the conventional decompilation tool, Ghidra, can be significantly improved by integrating it with LLMs.
Note that our approach aims at refining entire outputs from \ghidra{}, offering a broader strategy than merely recovering names or types~\cite{decompile_name,decompile_name2}.
We begin by detailing the general \derefine{} framework, and discuss our strategy to enhance Ghidra's output by LLM4Decompile-Ref.

\subsubsection{The Refined-Decompile Framework}

\begin{figure}[t]
  \centering
  \includegraphics[width=0.9\linewidth]{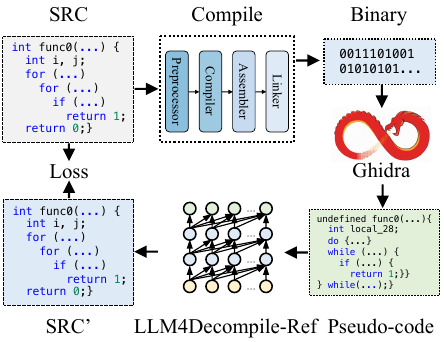}
  \caption{\derefine{} framework. It differs from \deend{} (Figure~\ref{fig:compile}) only in the LLM's input, which is pseudo-code decompiled from Ghidra.}
  \label{fig:refine}
\end{figure}
The \derefine{} approach is shown in Figure~\ref{fig:refine}. This approach differs from that in Figure~\ref{fig:compile} only in terms of the LLM's input, which in the case of \derefine{} comes from Ghidra's decompilation output. Specifically, Ghidra is used to decompile the binary, and then the LLM is fine-tuned to enhance Ghidra's output. While Ghidra produces high-level pseudo-code that may suffer from readability issues and syntax errors, it effectively preserves the underlying logic. Refining this pseudo-code significantly mitigates the challenges associated with understanding the obscure ASM.

\subsubsection{Refine Ghidra by LLM4Decompile-Ref}

\paragraph{\textbf{Decompiling using Ghidra.}} Decompiling the executable code with Ghidra (Figure~\ref{fig:refine}) is time-consuming due to the complex nature of the executables in ExeBench, which include numerous external functions and IO wrappers. Ghidra Headless~\cite{ghidraheadless} requires 2 seconds per sample using 128-core multiprocessing. Given such a high computational load, and the high similarities between non-executable and executable binaries, we choose to decompile the non-executable files using Ghidra. This choice significantly reduces the time to 0.2 seconds per sample, enabling us to efficiently gather large amounts of training data.

\paragraph{\textbf{Optimization Strategies.}} Similar to Section~\ref{sec:opt_end}, we augment our dataset by compiling with optimization levels O0, O1, O2, and O3. We further filter the dataset using LSH to remove duplicates. As shown in Figure~\ref{fig:case}, Ghidra often generates overly long pseudo-code. Consequently, we filter out any samples that exceed the maximum length accepted by our model.


\section{Experiments}
In this section, we discuss the experimental setups and results for LLM4Decompile-End and LLM4Decompile-Ref respectively.

\subsection{LLM4Decompile-End}
\subsubsection{Experimental Setups}
\paragraph{\textbf{Training Data.}}\label{sec:config}
As discussed in Section~\ref{sec:opt_end}, we construct asm-source pairs based on compilable and executable datasets from ExeBench~\cite{exebench}, where we only consider the decompilation of GCC~\cite{gcc_compile} compiled C function under x86 Linux platform. After filtering, our refined compilable training dataset includes 7.2 million samples, containing roughly 7 billion tokens. Our executable training dataset includes 1.6 million samples, containing roughly 572 million tokens.
To train the model, we use the following template: \texttt{\# This is the assembly code: [ASM code] \# What is the source code? [source code]}, where \texttt{[ASM code]} corresponds to the disassembled assembly code from the binary, and \texttt{[source code]} is the original C function. Note that the template choice does not impact the performance, since we fine-tune the model to produce the source code.

\paragraph{\textbf{Evaluation Benchmarks and Metrics.}}
To evaluate the models, we introduce HumanEval~\cite{humaneval} and ExeBench~\cite{exebench} benchmarks. HumanEval is the leading benchmark for code generation assessment and includes 164 programming challenges with accompanying Python solutions and assertions. We converted these Python solutions and assertions into C, making sure that they can be compiled with the GCC compiler using standard C libraries~\cite{gccstandard} and pass all the assertions, and name it HumanEval-Decompile. ExeBench consists of 5000 real-world C functions taken from GitHub with IO examples. 
Note that the HumanEval-Decompile consists of individual functions that depend only on the standard C library. In contrast, ExeBench includes functions extracted from real-world projects with user-defined structures and functions\footnote{Exebench provides comparison of the test set against the GitHub population using nine distinct code complexity metrics, confirming that the characteristics of the testing functions are aligned with functions in larger real-world projects.}.

As for the evaluation metrics, we follow previous work to calculate the re-executability rate~\cite{slade,refine_decompile}. During evaluation, the C source code is first compiled into a binary, then disassembled into assembly code, and fed into the decompilation system to be reconstructed back into C code. This decompiled C code is then combined with the assertions to check if it can successfully execute and pass those assertions. 

\paragraph{\textbf{Model Configurations.}}
The LLM4Decompile uses the same architecture as DeepSeek-Coder~\cite{deepseekcoder} and we initialize our models with the corresponding DeepSeek-Coder checkpoints. We employ Sequence-to-sequence prediction (S2S), which is the training objective adopted in most neural machine translation models that aim to predict the output given the input sequence. 
As illustrated in Equation \ref{eq2}, it minimizes the negative log likelihood for the source code tokens $x_i,...,x_j$:
\begin{equation}
    \mathcal L = -\sum_i \log P_i (x_i,...,x_j|x_1,...,x_{i-1}; \theta)
    \label{eq2}
\end{equation}
Where the loss is calculated only for the output sequence $x_i...x_j$, or the source code.

\begin{table*}[ht]
\centering

\begin{adjustbox}{width=1.95\columnwidth}

\begin{tabular}{ccccccccccc}
\toprule
\multirow{2}{*}{\textbf{Model/Benchmark}}      & \multicolumn{5}{c}{HumanEval-Decompile}                         & \multicolumn{5}{c}{ExeBench}                                                            \\
                       \cmidrule(lr){2-6}                                                                        \cmidrule(lr){7-11}
                       & O0              & O1              & O2              & O3              & AVG             & O0              & O1              & O2              & O3              & AVG             \\
\midrule
DeepSeek-Coder-6.7B    & 0.00          & 0.00          & 0.00          & 0.00          & 0.00          & 0.00          & 0.00          & 0.00          & 0.00          & 0.00          \\
GPT-4o                 & 30.49          & 11.59          & 10.37          & 11.59          & 16.01          & 4.43          & 3.28          & 3.97          & 3.43          & 3.78          \\
\hline
LLM4Decompile-End-1.3B & 47.20          & 20.61          & 21.22          & 20.24          & 27.32          & 17.86          & 13.62          & 13.20          & 13.28          & 14.49          \\
LLM4Decompile-End-6.7B & \textbf{68.05} & \textbf{39.51} & \textbf{36.71} & \textbf{37.20} & \textbf{45.37} & \textbf{22.89} & \textbf{16.60} & \textbf{16.18} & \textbf{16.25} & \textbf{17.98} \\
LLM4Decompile-End-33B  & 51.68          & 25.56          & 24.15          & 24.75          & 31.54          & 18.86          & 14.65          & 13.96          & 14.11          & 15.40          \\
\bottomrule
\end{tabular}

\end{adjustbox}
\caption{Main comparison of \deend{} approaches for re-executability rates on evaluation benchmarks.}
\label{table:main_results}
\end{table*}

\begin{table*}[h]
\centering

\begin{adjustbox}{width=1.95\columnwidth}

\begin{tabular}{ccccccccccc}
\toprule
\multirow{2}{*}{\textbf{Model/Benchmark}}     & \multicolumn{5}{c}{HumanEval-Decompile}                      & \multicolumn{5}{c}{ExeBench}                                                            \\ 
                                              \cmidrule(lr){2-6}                                             \cmidrule(lr){7-11}
                   & O0              & O1              & O2              & O3              & AVG             & O0              & O1              & O2              & O3              & AVG             \\ 
\midrule
Compilable-1.3B    & 42.68          & 16.46          & 16.46          & 17.07          & 23.17          & 5.68          & 4.46          & 4.16          & 4.43          & 4.68          \\
Compilable-6.7B    & \textbf{51.83} & \textbf{33.54} & \textbf{32.32} & \textbf{32.32} & \textbf{37.50} & 7.52          & 6.49          & 6.71          & 6.60          & 6.83          \\ \hline
Executable-1.3B    & 19.51          & 12.80          & 12.80          & 11.59          & 14.18          & 21.94          & 19.46          & 19.31          & 19.50          & 20.05          \\
Executable-6.7B    & 37.20          & 18.29          & 22.56          & 17.07          & 23.78          & \textbf{29.38} & \textbf{25.98} & \textbf{25.91} & \textbf{25.49} & \textbf{26.69} \\
\bottomrule
\end{tabular}

\end{adjustbox}
\caption{Ablation study on training dataset. The ``Compilable'' models are trained on 7.2M non-executable functions, while the ``Executable'' models are trained on 1.6M executable functions.}
\label{table:ablation_train}
\end{table*}

\paragraph{\textbf{Baselines.}}
We selected two key baselines for comparison. First, GPT-4o~\cite{gpt4} represents the most capable LLMs, providing an upper bound on LLM performance. 
Second, DeepSeek-Coder~\cite{deepseekcoder} is selected as the current SOTA open-source Code LLM. It represents the forefront of publicly available models specifically tailored for coding tasks. While recent work Slade~\cite{slade} fine-tunes language model for decompilation, it relies on intermediate compiler outputs, specifically, the *.s files. In practice, however, such intermediate files are rarely released by developers. Therefore, we focus on a more realistic approach, and consider decompilation only from the binaries, for further discussions please refer to Appendix~\ref{sec:appen_exe}.

\paragraph{\textbf{Implementation.}}\label{sec:implement}
We use the DeepSeek-Coder models obtained from Hugging Face~\cite{huggingface}. We train our models using LLaMA-Factory library~\cite{zheng2024llamafactory}. For 1.3B and 6.7B models, we set a $batch\:size=2048$ and $learning\;rate=2e{-}5$ and train the models for 2 epochs (15B tokens). Experiments are performed on NVIDIA A100-80GB GPU clusters. Fine-tuning the 1.3B and 6.7B LLM4Decompile-End takes 12 and 61 days on $8 \times A100$ respectively. Limited by the resources, for the 33B model we only train for 200M tokens. For evaluation, we use the \textit{vllm}~\cite{kwon2023efficient} to accelerate the generation (decompilation) process. We employ greedy decoding to minimize randomness.

\subsubsection{Experimental Results}
\paragraph{\textbf{Main Results.}}\label{sec:results}
Table~\ref{table:main_results} presents the re-executability rate under different optimization states for our studied models. The base version of DeepSeek-Coder-33B is unable to accurately decompile binaries. It could generate code that seemed correct but failed to retain the original program semantics. GPT-4o shows notable decompilation skills; it's capable to decompile non-optimized (O0) code with a success rate of 30.5\%, though the rate significantly decreases to about 11\% for optimized codes (O1-O3).
The LLM4Decompile-End models, on the other hand, demonstrate excellent decompilation abilities. The 1.3B version successfully decompiles and retains the program semantics in 27.3\% of cases on average, whereas the 6.7B version has a success rate of 45.4\%. This improvement underscores the advantages of using larger models to capture a program's semantics more effectively.
While attempting to fine-tune the 33B model, we encountered substantial challenges related to the high communication loads, which significantly slowed the training process and restricted us to using only 200M tokens (Section~\ref{sec:implement}). Despite this limitation, the 33B model still outperforms the 1.3B model, reaffirming the importance of scaling up the model size.

\setlength{\dashlinedash}{0.5pt}  
\setlength{\dashlinegap}{1.5pt}   
\setlength{\arrayrulewidth}{0.3pt}  

\begin{table*}[t]
\centering
\begin{adjustbox}{width=1.95\columnwidth}

\begin{tabular}{lcccccccccc}
\toprule
\multirow{2}{*}{\textbf{Model/Metrics}}               & \multicolumn{5}{c}{Re-executability Rate}                                               & \multicolumn{5}{c}{Edit Similarity}                                                     \\ 
\cmidrule(lr){2-6} \cmidrule(lr){7-11}
                  & O0              & O1              & O2              & O3              & AVG             & O0              & O1              & O2              & O3              & AVG             \\ 
\midrule
LLM4Decompile-End-6.7B  & 68.05          & 39.51          & 36.71          & 37.20          & 45.37          & 15.57          & 12.92          & 12.93          & 12.69          & 13.53          \\
\hdashline
\multicolumn{1}{l}{\ghidra{}} \\
\hspace{1em}Base                  & 34.76          & 16.46          & 15.24          & 14.02          & 20.12          & 6.99          & 6.13          & 6.19          & 5.47          & 6.20          \\
\hspace{1em}+GPT-4o                 & 46.95          & 34.15          & 28.66          & 31.10          & 35.22          & 6.60          & 5.63          & 5.67          & 4.99          & 5.72          \\
\hspace{1em}+LLM4Decompile-Ref-1.3B & 68.90          & 37.20          & 40.85          & 37.20          & 46.04          & 15.17          & 13.25          & 12.92          & 12.67          & 13.50          \\
\hspace{1em}+LLM4Decompile-Ref-6.7B & 74.39 & 46.95          & 47.56 & 42.07 & 52.74 & \textbf{15.59} & 13.53          & 13.42          & 12.73          & 13.82          \\
\hspace{1em}+LLM4Decompile-Ref-33B*  & 70.73          & 47.56 & 43.90          & 41.46          & 50.91          & 15.40          & \textbf{13.79} & \textbf{13.63} & \textbf{13.07} & \textbf{13.97} \\
\hspace{1em}+LLM4Decompile-Ref-22B*  & \textbf{80.49}          & \textbf{58.54} & \textbf{59.76}          & \textbf{57.93}          & \textbf{64.18}          & 15.19          & \textbf{14.04} & 13.58 & \textbf{13.40} & 13.85 \\
\bottomrule
\end{tabular}

\end{adjustbox}
\caption{Main comparison of \derefine{} approaches for re-executability rate and Edit Similarity on HumanEval-Decompile benchmark. ``+GPT-4o'' refers to enhance the Ghidra results with GPT-4o, ``+LLM4Decompile-Ref'' means refining Ghidra results with the fine-tuned LLM4Decompile-Ref models. Note that the 33B model was trained using only 200M tokens, which is just 10\% of the tokens used for the 1.3B/6.7B/22B model. For the 22B model, Please refer to Appendix~\ref{sec:baseline_models}.}
\label{table:bleu_run}
\end{table*}

\paragraph{\textbf{Ablation Study.}}\label{sec:ablation}
As discussed in Section~\ref{sec:config}, our training data comprises two distinct sets: 7.2 million compilable functions (non-executable) and 1.6M executable functions. We conducted an ablation study using these datasets, and the results are displayed in Table~\ref{table:ablation_train}. Here, ``Compilable'' denotes the model trained solely on compilable data, while ``Executable'' indicates models trained exclusively on executable data. Notably, the binary object from compilable functions lacks links to function calls, which is similar in text distribution to the HumanEval-Decompile data, consisting of single functions dependent only on standard C libraries. Consequently, the 6.7B model trained only on compilable data successfully decompiled 37.5\% of HumanEval-Decompile functions, but only 6.8\% on ExeBench, which features real functions with extensive user-defined functions. On the other hand, the 6.7B model trained solely on executable data achieved a 26.7\% re-executability rate on the ExeBench test set but faced challenges with single functions, with only a 23.8\% success rate on HumanEval-Decompile due to the smaller size of the training corpus. Limited by the space, we present further analysis in Appendix~\ref{sec:appen_error}.

\begin{figure}[h]
  \centering
  \includegraphics[width=1.04\linewidth]{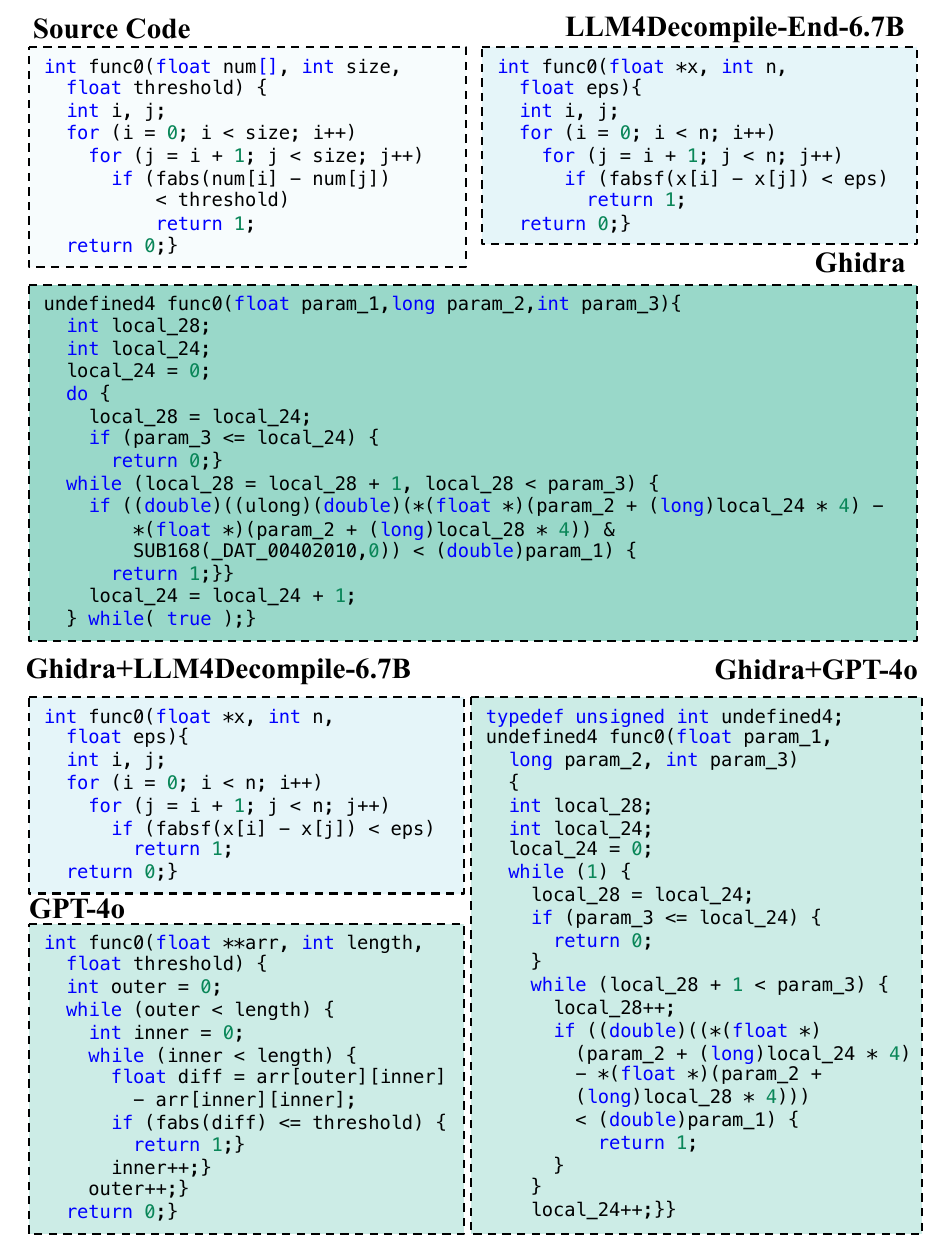}
  \caption{Decompilation results of different approaches. GPT-4o output is plausible yet fail to recover the array dimension (incorrect 2D array \texttt{arr[outer][inner]}). Ghidra's pseudo-code is notably less readable as discussed in Figure~\ref{fig:case}. GPT-refined Ghidra result (Ghidra+GPT-4o) marginally enhances readability but fails to correctly render \texttt{for} loops and array indexing. Conversely, LLM4Decompile-End and LLM4Decompile-Ref produce accurate and easy-to-read outputs. }
  \label{fig:read_case}
\end{figure}
\subsection{LLM4Decompile-Ref}
\subsubsection{Experimental Setups}
\paragraph{\textbf{Experimental Datasets.}}\label{sec:refine-data}
The training data is constructed using ExeBench, with Ghidra Headless~\cite{ghidraheadless} employed to decompile the binary object file. Due to constraints in computational resources, only 400K functions each with optimization levels from O0 to O3 (1.6M samples, 1B tokens) are used for training and the evaluation is conducted on HumanEval-Decompile. The models are trained using the same template described in Section~\ref{sec:config}. In addition, following previous work~\cite{btc,slade}, we access the readability of decompiled results in terms of Edit Similarity score.

\paragraph{\textbf{Implementation.}}
Configuration settings for the model are consistent with those in Section~\ref{sec:config}. For the 1.3B and 6.7B models, the fine-tuning process involves \textbf{2B tokens} in 2 epochs, and requires 2 and 8 days respectively on $8 \times A100$. Limited by the resource, for 33B model we only train for \textbf{200M tokens}. For evaluation, we first access the re-executability rate of Ghidra to establish a baseline. Subsequently, GPT-4o is used to enhance Ghidra's decompilation result with the prompt, \texttt{Generate linux compilable C/C++ code of the main and other functions in the supplied snippet without using goto, fix any missing headers. Do not explain anything.}, following DecGPT~\cite{refine_decompile}. Finally, we use LLM4Decompile-Ref models to refine the Ghidra's output.

\subsubsection{Experimental Results}

The results for the baselines and \derefine{} approaches are summarized in Table~\ref{table:bleu_run}. For the pseudo-code decompiled by Ghidra, which is not optimized for re-execution, only an average of 20.1\% of them pass the test cases. GPT-4o assists in refining this pseudo-code and enhancing its quality. The LLM4Decompile-Ref models offer substantial improvements over Ghidra's outputs, with the 6.7B model yielding a 160\% increase in re-executability. Similar to the discussion in Section~\ref{sec:results}, the 33B model outperforms the 1.3B model even though it used considerably less training data. And it achieves performance that is only 3.6\% below the 6.7B model, which benefited from ten times more training data. When compared to LLM4Decompile-End-6.7B, the LLM4Decompile-Ref-6.7B model, though trained on just 10\% of the data in LLM4Decompile-Ref models, shows a 16.2\% performance increase, suggesting a greater potential for the \derefine{} approach. We present further analysis in Appendix~\ref{sec:appen_data}.

An analysis of readability across different methods is also conducted and presented in Table~\ref{table:bleu_run} with illustrative examples presented in Figure~\ref{fig:read_case}. For text similarity, all decompiled outputs diverge from the original source code, with Edit Similarity ranging from 5.7\% to 14.0\%, primarily because the compilation process removes variable names and optimizes the logic structure. Ghidra generates pseudo-code that is particularly less readable with 6.2\% Edit Similarity on average. Interestingly, with refinement from GPT (Ghidra+GPT-4o), there is a marginal decrease in Edit Similarity. GPT assists in refining type errors like \texttt{undefined4} and \texttt{ulong} (Figure~\ref{fig:read_case}). However, it struggles to accurately reconstruct \texttt{for} loops and array indexing. In contrast, both LLM4Decompile-End and LLM4Decompile-Ref generate outputs that are more aligned with the format of the source code and easier to comprehend. To summarize, domain-specific fine-tuning is crucial for enhancing re-executability and readability of decompilation outputs.

We further employed GPT-4o to evaluate readability~\cite{wang-etal-2023-chatgpt,liu-etal-2023-g}. Specifically, we guide GPT to assess syntax similarity (variables, loops, conditions) and structural integrity (logic flow, structure) using a structured template. We then summarize readability with a score from 1 (Poor) to 5 (Excellent), based on detailed comparisons between original and decompiled code. The template is available on our GitHub repository\footnote{\url{https://github.com/albertan017/LLM4Decompile/blob/main/samples/readability_template.txt}}. Table~\ref{table:readability} summarizes our readability assessments on HumanEval-Decompile across various models and optimization levels.

\begin{table}[h]
\centering

\begin{adjustbox}{width=0.95\columnwidth}

\begin{tabular}{cccccc}
\hline
Optimization Level     & O0     & O1     & O2     & O3     & AVG    \\ \hline
GPT-4o                 & 2.8171 & 2.3537 & 2.2927 & 2.311  & 2.4436 \\
Ghidra                 & 2.9756 & 2.4085 & 2.5183 & 2.3841 & 2.5716 \\
LLM4Decompile-End-6.7B & 4.0732 & 3.4634 & 3.4024 & 3.2378 & 3.5442 \\ \hline
\end{tabular}

\end{adjustbox}
\caption{Evaluation by GPT-4o on the readability of decompiled results from various methods.}
\label{table:readability}
\end{table}
Compared with the results in Table~\ref{table:bleu_run}, it indicates that Edit Similarity (ES) follows a trend similar to GPT evaluation. Although ES is mathematically based, its values can be difficult to interpret. For instance, a 15 ES score obtained by LLM4Decompile model may seem low, yet the decompiled function and the source code are highly aligned. In contrast, GPT evaluation, which measures readability conceptually, is more intuitive. A score of 4 on the GPT scale suggests that the decompiled code is nearly identical to the original. Nonetheless, these scores are derived from GPT's "subjective" judgments. Combining insights from both ES and GPT-Eval could lead to a more thorough assessment of code readability.




\begin{table*}[h]
\centering
\begin{adjustbox}{width=1.95\columnwidth}

\begin{tabular}{ccccccccccc}
\toprule
\multirow{2}{*}{\textbf{Model/Obfuscation}}       & \multicolumn{5}{c}{Control Flow Flattening} & \multicolumn{5}{c}{Bogus Control Flow}     \\
\cmidrule(lr){2-6}                                                                        \cmidrule(lr){7-11}
                        & O0      & O1     & O2     & O3     & AVG    & O0     & O1     & O2     & O3     & AVG    \\ 
\midrule
LLM4Decompile-End-6.7B  & 4.27  & 4.88 & 4.88 & 3.05 & 4.27 & 9.76 & 7.32 & 7.93 & 9.76 & 8.69 \\
Ghidra                  & 12.20  & 6.71 & 6.10 & 6.71 & 7.93 & 6.10 & 4.27 & 3.05 & 4.27 & 4.42 \\
+LLM4Decompile-Ref-6.7B & 6.71  & 3.66 & 4.88 & 5.49 & 5.19 & 15.85 & 14.02 & 8.54 & 7.93 & 11.59 \\ 
\bottomrule
\end{tabular}

\end{adjustbox}
\caption{Re-executability rates of different approaches on the HumanEval-Decompile benchmark under obfuscations. Compared to Table~\ref{table:bleu_run}, the decompilation success rates significantly drop for over 70\%.}
\label{table:obfuscation}
\end{table*}

\section{Obfuscation Discussion}\label{sec:obf}
The process of decompilation aims at revealing the source code from binaries distributed by developers, presenting a potential threat to the protection of intellectual property. To resolve the ethical concerns, this section accesses the risks of the possible misuse of our decompilation models. 

In software development, engineers typically implement obfuscation techniques before releasing binary files to the public~\cite{dobf-cite,ollvm}. This is done to protect the software from unauthorized analysis or modification. In our study, we focus on two fundamental obfuscation techniques as suggested in Obfuscator-LLVM~\cite{ollvm}: Control Flow Flattening (CFF) and Bogus Control Flow (BCF). These techniques are designed to disguise the true logic of the software, thereby making decompilation more challenging to protect the software's intellectual property. We present the details of these two techniques in the Appendix~\ref{sec:appen_obf}.

Results summarized in Table~\ref{table:obfuscation} demonstrate that basic conventional obfuscation techniques are sufficient to prevent both Ghidra and LLM4Decompile from decoding obfuscated binaries.  For example, the decompilation success rate for the most advanced model, LLM4Decompile-Ref-6.7B, drops significantly for 90.2\% (0.5274 to 0.0519) under CFF and 78.0\% (0.5274 to 0.1159) under BCF. Considering the industry standard of employing several complex obfuscation methods prior to software release, experimental results in Table~\ref{table:obfuscation} mitigate the concerns about unauthorized use for infringement of intellectual property.

\section{Conclusions}
We propose LLM4Decompile, the first and largest open-source LLM series with sizes ranging from 1.3B to 33B trained to decompile binary code.
Based on the \deend{} approach, we optimize the LLM training process and introduce the LLM4Decompile-End models to decompile binary directly. The resulting 6.7B model shows a decompilation accuracy of 45.4\% on HumanEval and 18.0\% on ExeBench, surpassing existing tools like Ghidra and GPT-4o over 100\%.
Additionally, we improve the \derefine{} strategy to fine-tune the LLM4Decompile-Ref models, which excel at refining the Ghidra's output, with 16.2\% improvement over LLM4Decompile-End. Finally, we conduct obfuscation experiments and address concerns regarding the misuse of LLM4Decompile models for infringement of intellectual property.

\section*{Limitations}
The scope of this research is limited to the compilation and decompilation of C language targeting the x86 platform. While we are confident that the methodologies developed here could be easily adapted to other programming languages and platforms, these potential extensions have been reserved for future investigation. 
Furthermore, Our research is limited by financial constraints, with a budget equivalent to using $8 \times A100$ GPUs for one year, which includes all trials and iterations. As a result, we have only managed to fully fine-tune models up to 6.7B, and conducted initial explorations on the 33B models with a small dataset, leaving the exploration of 70B and larger models to future studies. Nonetheless, our preliminary tests confirm the potential advantages of scaling up model sizes and suggest a promising direction for future decompilation research into larger models.

\section*{Ethic Statement}

We have evaluated the risks of the possible misuse of our decompilation models in Section~\ref{sec:obf}. Basic obfuscation methods such as Control Flow Flattening and Bogus Control Flow have been empirically tested and proven to protect against unauthorized decompilation by both traditional tools like Ghidra and advanced models like LLM4Decompile. This built-in limitation ensures that while LLM4Decompile is a powerful tool for legitimate uses, it does not facilitate the infringement of intellectual property.

In practical applications in the industry, software developers typically employ a series of complex obfuscation methods before releasing their software. This practice adds an additional layer of security and intellectual property protection against decompilation. LLM4Decompile's design and intended use respect these measures, ensuring that it serves as an aid in legal and ethical scenarios, such as understanding legacy code or enhancing cybersecurity defenses, rather than undermining them.

The development and deployment of LLM4Decompile are guided by strict ethical standards. The model is primarily intended for use in scenarios where permission has been granted or where the software is not protected by copyright. This includes academic research, debugging, learning, and situations where companies seek to recover lost source code of their own software.

\section*{Acknowledgments}
This work is partially supported by the National Natural Science Foundation of China (No. 62372220), the Research Grants Council of the Hong Kong Special Administrative Region, China (Project No. PolyU/25200821), the NSFC Young Scientists Fund (Project No. 62006203), the Innovation and Technology Fund (Project No. PRP/047/22FX), and PolyU Internal Fund from RC-DSAI (Project No. 1-CE1E).

\bibliography{custom}

\appendix


\section{ExeBench Setups}\label{sec:appen_exe}
For every sample in ExeBench's executable splits, assembly code from *.s file—a compiler's intermediate output as discussed in Section~\ref{sec:end} and Figure~\ref{fig:case}—is required to compile the sample into a binary. The specific compilation settings and processing details, however, are not provided by their authors. Consequently, we choose to compile the code in a standard way and manage to compile only half of the samples. This leaves us with 443K out of 797K samples for the executable training set and 2621 out of 5000 samples for the executable test set. Accordingly, we train our model on the 443K samples and conduct the re-executability evaluation on these 2621 samples, the results are shown in Table~\ref{table:main_results}.

The researchers from Slade~\cite{slade}, who also developed ExeBench~\cite{exebench}, have published their decompilation findings on ExeBench. They chose to decompile the intermediate output, or assembly code from *.s file, directly without further compilation into binaries, where in practice, such intermediate output is rarely released by software developers. Their reported results, as seen in Table~\ref{table:appen}, show a significant difference from ours. 
Their version of ChatGPT achieved a re-executability rate of 22.2\% and an edit similarity of 44.0\% under O0 optimization. On the other hand, our GPT-4o model only reached a 4.4\% re-executability rate and a 7.9\% edit similarity. The approach taken by Slade involves settings not commonly available in practical decompilation scenarios, which explains why their results vary significantly from ours. We adhere to a more realistic setting, decompiling binary files based solely on their intrinsic data, without any external information.

\begin{table}[h]
\centering

\begin{adjustbox}{width=0.95\columnwidth}

\begin{tabular}{ccccc}
\hline
Model/Metrics      & \multicolumn{2}{c}{Re-executability} & \multicolumn{2}{c}{Edit Similarity} \\ \hline
Optimization Level & O0                & O3               & O0               & O3               \\ \hline
Slade              & 59.5              & 52.2             & 71.0             & 60.0             \\
ChatGPT            & 22.2              & 13.6             & 44.0             & 34.0             \\
GPT-4o(ours)       & 4.4               & 3.4              & 7.9              & 6.6              \\ \hline
\end{tabular}

\end{adjustbox}
\caption{Re-executability and Edit Similarity on Exebench.}
\label{table:appen}
\end{table}

\begin{figure}[t]
  \centering
  \includegraphics[width=0.95\linewidth]{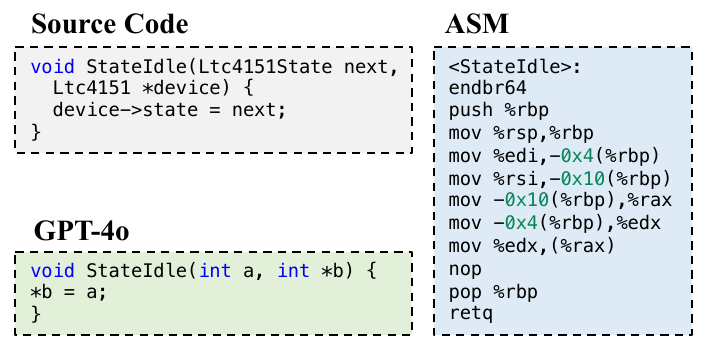}
  \caption{Decompilation results of GPT-4o on ExeBench test case.}
  \label{fig:appen}
\end{figure}

To further illustrate our settings, Figure~\ref{fig:appen} offers an example where the source function includes specific user-defined types like \texttt{Ltc4151State}, \texttt{Ltc4151}, and \texttt{device}. However, these types are completely lost after compilation, i.e., no information related to these user-definitions can be found in the binary (disassembled ASM code). Consequently, GPT-4o is unable to reconstruct these types based purely on the ASM (the realistic setting), but converting them to default types \texttt{int} or \texttt{pointer}, producing non-executable code. This issue was pervasive across the ExeBench test set, leading to the failure of GPT-4o models in decompiling the ExeBench samples in a realistic setting.

\section{Compilable and Executable Binary}\label{sec:appen_exe}
The statistics of training and testing datasets is summarized in the Table~\ref{table:length}. We also present an example to illustrate the difference between these two datasets.
\begin{table}[h]
\centering

\begin{adjustbox}{width=0.6\columnwidth}

\begin{tabular}{ccc}
\hline
Dataset/Code         & ASM    & SRC    \\ \hline
Train-Executable     & 205.08 & 119.35 \\
Test-Exebench        & 280.27 & 162.68 \\
Train-Compilable     & 711.89 & 241.16 \\
Test-Decompile-Eval  & 808.07 & 186.84 \\ \hline
\end{tabular}

\end{adjustbox}
\caption{Statistics of training and testing set.}
\label{table:length}
\end{table}

\begin{figure}[h]
  \centering
  \includegraphics[width=0.85\linewidth]{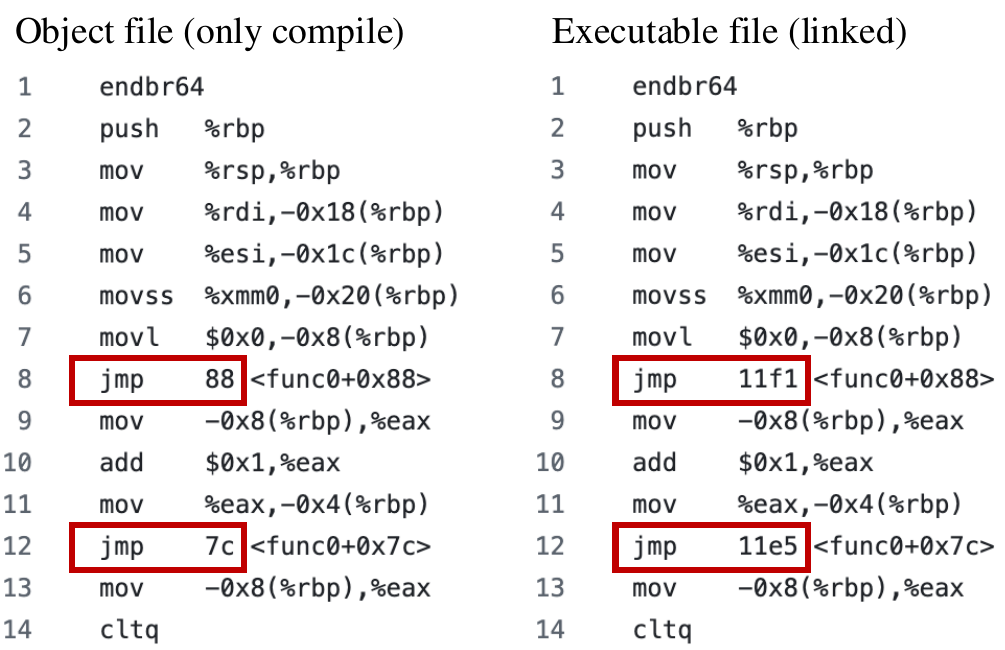}
  \caption{Compilable data and Executable data.}
  \label{fig:comp_exe}
\end{figure}

As shown in Figure~\ref{fig:comp_exe}, the primary distinction between a compilable binary and an executable binary is the handling of function operation addresses. In a compilable file, the address for a jump operation is placeholder, representing only a relative offset within the function. Conversely, in an executable file, this jump operation address is resolved during the linking process, pointing directly to the specific memory location where the code will execute.

\section{Further Analysis of LLM4Decompile-End}\label{sec:appen_error}
\begin{figure}[h]
  \centering
  \includegraphics[width=0.8\linewidth]{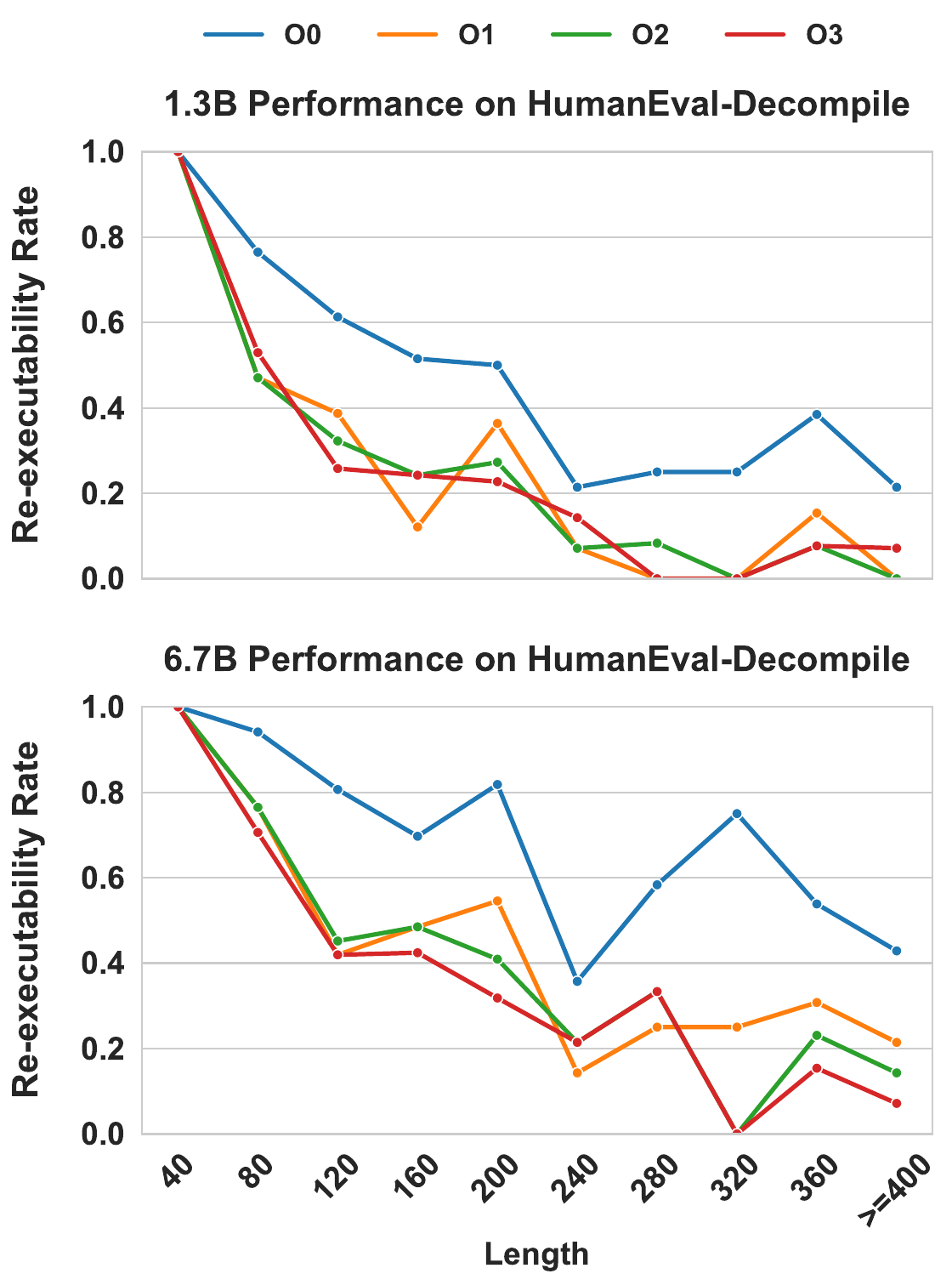}
  \caption{Re-executability rate with the growth of input length. The 6.7B model is more robust against input length.}
  \label{fig:length}
\end{figure}

\begin{figure}[h]
  \centering
  \includegraphics[width=0.85\linewidth]{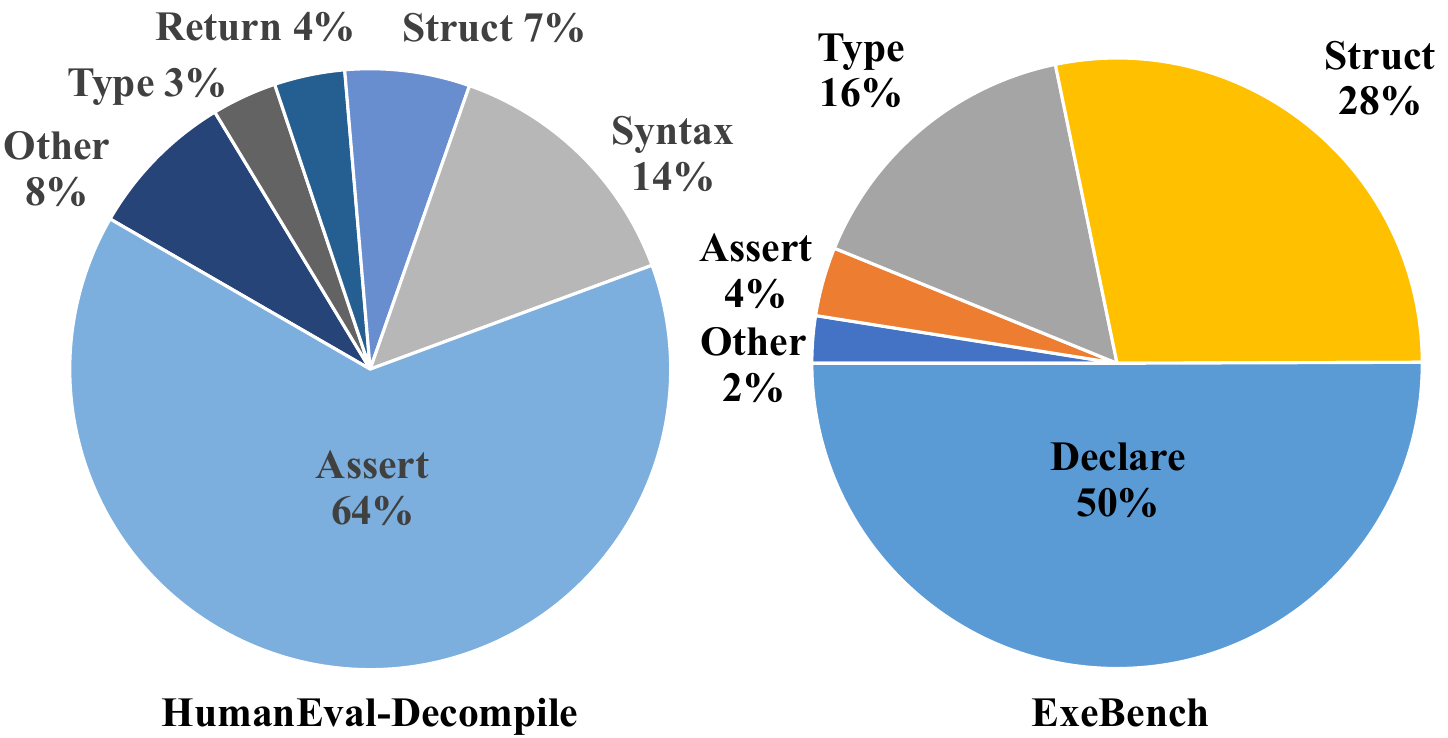}
  \caption{Types of errors identified in the two benchmarks: LLM4Decomile-End-6.7B faces issues with logical errors in HumanEval-Decompile and user-defined components in ExeBench. }
  \label{fig:error}
\end{figure}
Figure~\ref{fig:length} illustrates that the re-executability rate decreases as the input length increases, and there is a marked decline in performance at higher levels of code optimization, highlighting the difficulties in decompiling long and highly optimized sequences. Importantly, the performance difference between the 1.3B and 6.7B models showcased in the figure emphasizes the advantages of larger models in such tasks. Larger models, with their expanded computational resources and deeper learning capabilities, are inherently better at resolving the challenges posed by complex decompilations.

The error analysis presented in Figure~\ref{fig:error} for LLM4Decompile-End-6.7B indicates that logical errors are prevalent in the HumanEval-Decompile scenarios, with 64\% of errors due to assertions that the decompiled codes do not pass. In the ExeBench dataset, which features real functions with user-defined structures and types, the major challenges are related to reclaiming these user-specific components. Where 50\% of the errors come from undeclared functions, and 28\% from improper use of structures. Given that these user-defined details are typically lost during the compilation process, reconstructing them can be particularly challenging. Integrating techniques like Retrieval Augmented Generation~\cite{meta2020rag} might supplement the decompilation process with necessary external information.

\section{Data Quality, Volume and Model}\label{sec:appen_data}
\paragraph{Data Quality}
In this project, we intentionally limited our data preprocessing to classical techniques such as filtering short texts and removing duplicates. This approach was chosen to establish a fair baseline model for decompilation that minimizes potential biases, aiming to provide a broad, unrefined baseline model that reflects diverse scenarios. We acknowledge that selective data removal, specifically excluding data incompatible with standard C libraries~\cite{gccstandard}, can enhance performance, as evidenced in Table~\ref{table:standard_c} with Decompile-Eval, which only relies on standard C libraries. While refining the dataset can lead to improved performance, our primary goal in this study was to set a foundational baseline for the community. We believe this baseline can serve as a starting point encouraging future research to refine and expand.

\begin{table}[h]
\centering

\begin{adjustbox}{width=0.95\columnwidth}

\begin{tabular}{cccccc}
\hline
Models                           & \multicolumn{5}{c}{Re-executability Rate} \\ \hline
Optimization Level               & O0                    & O1     & O2     & O3     & AVG    \\ \hline
Compilable-6.7B                  & 51.83                & 33.54 & 32.32 & 32.32 & 37.50 \\
+Executable (2B tokens)          & 68.05                & 39.51 & 36.71 & 37.20 & 45.37 \\
+Exe w. Standard C (100M tokens) & 71.80                & 42.68 & 41.31 & 41.46 & 49.31 \\ \hline
\end{tabular}

\end{adjustbox}
\caption{Performance improves when training data (+Exe w. Standard C) closely resembles the testing set patterns.}
\label{table:standard_c}
\end{table}

\paragraph{Data Volume}
We have summarized the relationship between performance and training epochs (2B token in one epoch) for the LLM4Decompile-Ref-1.3B model in Table~\ref{table:1b4epoch}. From this, it is clear that a single epoch serves as a strong baseline, while two epochs optimize performance. Additional epochs tend to lead to overfitting and diminished results.
\begin{table}[h]
\centering

\begin{adjustbox}{width=0.7\columnwidth}

\begin{tabular}{cccccc}
\hline
Epoch & O0     & O1     & O2     & O3     & Avg    \\ \hline
1     & 67.68 & 41.46 & 41.46 & 35.37 & 46.49 \\
2     & 68.29 & 40.85 & 40.85 & 37.20 & 46.80 \\
3     & 62.80 & 37.80 & 36.59 & 29.88 & 41.77 \\
4     & 51.22 & 32.93 & 27.44 & 26.22 & 34.45 \\ \hline
\end{tabular}

\end{adjustbox}
\caption{Performance of LLM4Decompile-Ref-1.3B on HumanEval-Decompile w.r.t. different training epochs.}
\label{table:1b4epoch}
\end{table}

Moreover, we have included results from the 6.7B model to address scaling issues and provided comparison with the 1.3B and 33B models in Table~\ref{table:modelepoch}. Notably, the 6.7B model achieves comparable performance to the 1.3B model with only 20\% of the data, and the 33B model reaches similar outcomes to the 6.7B with just 10\% of the data, although these ratios may differ with varying datasets.
\begin{table}[h]
\centering

\begin{adjustbox}{width=0.8\columnwidth}

\begin{tabular}{ccccccc}
\hline
Size & epoch & O0     & O1     & O2     & O3     & Avg    \\ \hline
1.3B & 1.0   & 67.68 & 41.46 & 41.46 & 35.37 & 46.49 \\
6.7B & 0.05  & 55.49 & 31.71 & 34.76 & 30.49 & 38.11 \\
6.7B & 0.1   & 57.93 & 36.74 & 32.77 & 32.01 & 39.86 \\
6.7B & 0.2   & 65.85 & 37.80 & 40.24 & 34.76 & 44.66 \\
6.7B & 0.5   & 65.55 & 45.73 & 43.29 & 43.75 & 49.58 \\
6.7B & 1.0   & 72.56 & 45.73 & 43.90 & 42.68 & 51.22 \\
33B  & 0.1   & 70.73 & 47.56 & 43.90 & 41.46 & 50.91 \\ \hline
\end{tabular}

\end{adjustbox}
\caption{Performance of LLM4Decompile-Ref models on HumanEval-Decompile w.r.t. different training epochs.}
\label{table:modelepoch}
\end{table}

Our findings suggest that model scaling can significantly enhance performance when the training data is adequately large (100M tokens), but repetitive training risks overfitting after a few epochs.

\paragraph{Model}\label{sec:baseline_models}
Choosing the right base model for decompilation training significantly influences performance. Our first choice, Deepseek-Coder-6.7B, delivered an encouraging average re-executability rate of 52.74\% on the HumanEval-Decompile benchmark. Conversely, LLM-Compiler-7B~\cite{llmcompiler}, trained to compile source code into LLVM IR~\cite{lattner2004llvm}—the opposite of decompilation—served as a more effective foundation, enhancing performance by 3.5\% compared to Deepseek-Coder-6.7B. Additionally, Yi-Coder-9B~\cite{yicoder}, introduced in September 2024 as the current state-of-the-art model, markedly improved decompilation training results by 23.1\%. Furthermore, CodeStral-22B~\cite{codestral}, benefiting from its larger architecture, provided a 21.7\% improvement over smaller models.

\begin{table}[h]
\centering

\begin{adjustbox}{width=0.95\columnwidth}

\begin{tabular}{cccccc}
\hline
Model/Opt. Level    & O0             & O1             & O2             & O3             & AVG            \\ \hline
DeepSeek-Coder-6.7B & 74.39          & 46.95          & 47.56          & 42.07          & 52.74          \\
LLM-Compiler-7B     & 72.56          & 51.83          & 48.78          & 45.12          & 54.57          \\
Yi-Coder-9B         & 79.27          & \textbf{62.20} & \textbf{61.59} & 56.71          & \textbf{64.94} \\
CodeStral-22B       & \textbf{80.49} & 58.54          & 59.76          & \textbf{57.93} & 64.18          \\ \hline
\end{tabular}

\end{adjustbox}
\caption{Comparison of re-executability rates for base models in the LLM4Decompile-Ref series.}
\label{table:basemodel}
\end{table}

\section{Obfuscation Techniques}\label{sec:appen_obf}
We provide the details of two classic obfuscation techniques suggested in Obfuscator-LLVM~\cite{ollvm}.
\paragraph{Control Flow Flattening.} It enhances the security of software by transforming its straightforward, hierarchical control flow into a more complex, flattened structure. The workflow involves breaking a function into basic blocks, arranging these blocks at the same level, and encapsulating them within a switch statement inside a loop. 

\paragraph{Bogus Control Flow.} It modifies a function's execution sequence by inserting an additional basic block prior to the existing one. This added block includes an opaque predicate, followed by a conditional jump that leads back to the original block. Additionally, the original basic block is polluted with randomly selected, meaningless instructions.

\end{document}